# Signal generation and storage
# in FRET-based nanocommunications

Jakub Kmiecik[*1], Krzysztof Wojcik[*2], Pawel Kulakowski[†1], Andrzej Jajszczyk[1]

[1]*Department of Telecommunications, AGH University of Science and Technology, Krakow, Poland*
[2]*2nd Department of Internal Medicine, Faculty of Medicine, Jagiellonian University Medical College, Krakow, Poland*

[*]*These authors contributed equally to this work*
[†]*Corresponding author: Pawel Kulakowski (kulakowski@kt.agh.edu.pl)*

## Abstract

The paper is concerned with Förster Resonance Energy Transfer (FRET) considered as a mechanism for communication between nanodevices. Two solved issues are reported in the paper, namely: signal generation and signal storage in FRET-based nanonetworks. First, luciferase molecules as FRET transmitters which are able to generate FRET signals themselves, taking energy from chemical reactions without any external light exposure, are proposed. Second, channelrhodopsins as FRET receivers, as they can convert FRET signals into voltage, are suggested. Further, medical in-body systems where both molecule types might be successfully applied, are discussed. Luciferase-channelrhodopsin communication is modeled and its performance is numerically validated, reporting on its throughput, bit error rate, propagation delay and energy consumption.

## 1. Introduction

With the current rapid development of technology, electronic and mechanical devices can be much smaller than even a decade ago, at the same time gaining new operating functionalities. However, when the device scale goes to the size of single molecules, cooperation between such devices and contacting them from outside their networks become quite challenging issues. Förster Resonance Energy Transfer (FRET), a phenomenon characterized by very small propagation delays and a good throughput, has already been proposed for nanocommunications [1] and has been proved to be an efficient mechanism for transferring data at nanodistances [2–4]. Research on FRET for nanocommunication purposes and its integration with other elements on nanoworld is though at an early stage.

The main contribution of the paper is solving two issues that, until now, were not solved for FRET-based communications [5]. The first issue is concerned with **FRET signal storage**. A molecule receiving a FRET signal cannot hold it and the signal must quickly be released, usually emitting a photon. Here, we propose to use channelrhodopsins which are molecules that can be used as nano-converters, changing FRET signals into a voltage which can be detected by other devices. The second issue concerns **FRET signal generation**. In FRET nanocommunication experiments so far, a FRET transmission was initiated by an external laser source, which is rather troublesome inside a human body. Instead, we propose bioluminescent luciferase molecules as FRET transmitters, which are able to generate FRET signals themselves, taking energy from chemical reactions. We further discuss medical in-body systems where such FRET-based communication between luciferases and channelrhodopsins might be applied. We consider a system of small vesicles, liposomes, circulating through human vascular system and collecting data about body condition. The data gathered by liposomes are delivered to a detector placed in a human vein. Data transmission is realized by a luciferase-channelrhodopsin FRET interface. We model layers of luciferase and channelrhodopsin molecules based on their molecular structures and then we use their spectral characteristics to assess the efficiency of the FRET communication. Finally, we numerically validate the communication performance, reporting the bit error rate, throughput, propagation delay and the energy consumption.

The rest of the paper is organized as follows. In Section 2, we review various mechanisms considered for nanocommunications. In Section 3, the FRET phenomenon is explained. We present luciferase and channelrhodopsin molecules in Section 4, discussing their properties and possible communication between them. In Section 5, we discuss medical applications where such a communication might be applied, focusing on data gathering systems based on liposomes. In Section 6, we provide a model for communication between luciferases and channelrhodopsins, while in Section 7 we numerically evaluate the communications performance. Finally, in Section 8, we conclude the paper, also discussing some open issues.

## 2. Molecular communication mechanisms

So far, there has been a number of molecular communication techniques proposed that can be categorized according to their propagation mechanisms. According to [6], the following ones can be distinguished: free diffusion based propagation, flow assisted propagation, active transport and bacteria assisted propagation. This classification can be supplemented with well investigated neural communication. Transmission parameters, such as delay, throughput and bit error rate (BER), differ for each of these propagation mechanisms.

Free diffusion based propagation systems usually consist of three main components: transmitter, receiver and signaling molecules that carry information. Signaling molecules freely diffuse over the liquid medium. In order to send bit '1' the transmitter releases a substantial number of molecules, while sending bit '0' is done by releasing significantly fewer number of molecules or not releasing them at all. The performance of this communication method depends on the number of factors such as: diffusion coefficient, the ratio of signaling molecules representing single bits, size and concentration of signaling molecules, rate of molecule release, transmitter–receiver distance and applied detection rules. The capacity was estimated to be in the range between 1 kb/s and 3 kb/s [7] and BER was reported as near-zero value for distances about few micrometers with energy consumption about $10^{-14} - 10^{-13}$ J per bit. Diffusion based propagation systems can be enhanced by a flow directed from the transmitter to the receiver increasing communication range and maintaining BER on the order of $10^{-5}$ [8].

In active transport molecular communication a cargo conveying information molecules is attached on a transmitter side to a polymer called molecular motor. The molecular motor is able to carry this cargo and literally walk over a cylindrical molecule called microtubule. The molecular motor walks along a microtubule until reaches a receiver. Such communication is characterized by a large delay – tens of seconds and limited range – up to ten micrometers [9]. Near-zero BER can only be achieved for a sufficiently long inter symbol duration time – more than a hundred seconds. A slightly different approach to active transport is presented in [10, 11]: molecular motors are fixed next to each other to a microtubule, connecting the transmitter and receiver sides. Another shorter microtubule carrying cargo containing information particles is attached to the molecular motors and propagates along them. The reported capacity for such a communication channel can reach up to 5 bit/s depending on the number of information particles.

In bacteria assisted propagation, a bacterium is an information carrier [12]. Information is stored in DNA strands that are enclosed in a flagellated bacterium which swims in a liquid environment or over the tissue surface. By using few hundreds of bacteria an error-free transmission over few millimeters can be achieved, but at the expense of huge delay, measured in hours.

Neurons communicate with each other by neurotransmitters. When a presynaptic nerve cell is stimulated, an electrical impulse called a spike can be initiated. After the spike traverses the presynaptic neuron's axon and arrives at the axonal terminal, it triggers the release of neurotransmitters that diffuse to postsynaptic nerve cell to stimulate it. Neural communication was already studied on the subject of the nerve stimulation frequency, distance between nerve cells [13], spike width and the number of axonal terminals [14]. It was shown that the neural communication channel capacity can be up to 3000 bit/s (assuming one bit per spike) and the BER can be at level $10^{-4}$ at a distance of few hundreds of micrometers, with energy consumption about $10^{-16} - 10^{-13}$ J per bit.

## 3. FRET phenomenon for nanocommunications

Förster Resonance Energy Transfer is a phenomenon where a signal may be passed between two molecules irradiatively [15]. Application of this phenomenon to nanocommunications has been already proposed and thoroughly investigated [1–3]. In FRET, there is an excited molecule, called donor, at the transmitter side, and another molecule, called acceptor, at the receiver side of the communication channel. If the emission spectrum of the donor matches the absorption spectrum of the acceptor, the excitation energy may pass, without any radiation, from the transmitter, i.e., the donor molecule, to the receiver, i.e., the acceptor molecule[1]. The donor and the acceptor must be located very close to each other, usually of few nanometers, so this is a short range communication mechanism. FRET has proved to have very small signal propagation delays, on the order of nanoseconds, and a throughput of several Mbit/s [4], and, therefore, it is very suitable for future networks of nanomachines when cooperating and exchanging information between each other. Recently, routing techniques for FRET-based nanonetworks have also been proposed [5]. The FRET phenomenon between a single donor and a single acceptor is not very efficient, as its efficiency decreases with the sixth power of the donor-acceptor distance $r$ [17]:

$$E = \frac{R_0^6}{r^6 + R_0^6}, \qquad (1)$$

where $R_0$ is so called Förster distance, which is a value characterizing the spectral match of donor emission and acceptor absorption spectra. Förster distance can be obtained experimentally or calculated when the donor and acceptor spectral characteristics are known [17]:

$$R_0^6 = 0.211 \cdot \kappa^2 n_r^{-4} Q_D \int F_D(\lambda) \varepsilon_A(\lambda) \lambda^4 d\lambda, \qquad (2)$$

where $F_D(\lambda)$ is the donor emission spectrum with the total intensity normalized to one and $\varepsilon_A(\lambda)$ is the acceptor molar extinction coefficient in the function of wavelength, which can be directly derived from the acceptor emission spectrum. The integral of the product of these two spectra is called the overlap integral. $Q_D$ is the quantum yield of the donor molecule in absence of the acceptor molecule – this value describes the ratio of emitted photons to absorbed photons, $n_r$ is the refractive index which is assumed to be 1.4 for biomolecules in aqueous solutions and $\kappa^2$ is a parameter describing a relative orientation of donor and acceptor dipoles which is assumed to be equal 0.476 [17], a value appropriate for the situation when the relative donor-acceptor orientation does not change during the lifetime of the excited state.

---

[1] In some cases, an energy transfer between molecules that are not matched spectrally is also possible; see the phenomenon called Phonon-Assisted Energy Transfer [16].

While FRET between single molecules is not efficient, it has been already shown that using multiple donors and acceptors, in the so called MIMO-FRET communications, FRET efficiency may be quite high and the transmission bit error rate below $10^{-3}$ may be easily reached [3, 4]. Transmission of '0' and '1' bits can be realized with the ON-OFF modulation [18]. While sending bit '0', donors are not excited, resulting in no energy transfer to acceptors. It is assumed here that such a transmission is always correct (no random excitations that could cause false detections), if only a proper synchronization is kept between the transmitter and the receiver. When bit '1' is going to be sent, all donors are excited at the same time. If at least one of them passes its excitation energy to any of the acceptors, the transmission is successful. With $n$ donors and $m$ acceptors, FRET efficiency is equal to [4]:

$$E_{n,m} = 1 - \left(1 - \frac{m \cdot R_0^6}{r^6 + m \cdot R_0^6}\right)^n \tag{3}$$

And the respective bit error rate is given by:

$$\text{BER}_{n,m} = 0.5 \cdot \left(\frac{r^6}{r^6 + m \cdot R_0^6}\right)^n \tag{4}$$

## 4. Luciferases and channelrhodopsins

The main contribution of this paper is the proposal of layers of luciferases, at the transmitter side, and channelrhodopsins, at the receiver side, for FRET-based nanocommunications. Usage of luciferases may solve the problem of FRET signal generation, as in this case no laser is needed. Instead, the signal transmission may be initiated by a chemical reaction. On the other hand, application of channelrhodopsins will help with the FRET signal storage. Channelrhodopsins are able to convert FRET signals into the electrical voltage.

### 4.1. Luciferases

A FRET transmitter, i.e., a donor, in most cases is excited externally, e.g., via a laser beam. In an in-body medical system, such an approach is yet questionable, as it is the transmitting molecule itself which should be responsible for initiating the data transmission. In this situation, instead of typical FRET donors, it is much better to use bioluminescent molecules [19, 20]. For bioluminescent donors, the source of excitation energy is not external; instead it lies in a biochemical reaction occurring close to such a molecule[2]. Thus, the transmitter may be excited by the transmitting nanomachine itself, and after that, the signal passes to the receiver in the same manner as in typical FRET transmission, i.e., without any radiation.

The bioluminescent donors are, e.g., luciferase molecules [21]. A luciferase is an enzyme that converts chemical energy to light or FRET signals requiring two other molecule types for the reaction: a) luciferins (here, we used furimazine molecules which are a kind of luciferins, 1.5 nm in diameter), and b) oxygen [22]. The main drawback of luciferases is their quite long time of relaxation, as, e.g., NanoLuc luciferase reaction turnover is ca. 3.8 $s^{-1}$, assuming the ambient temperature about 36 degrees Celsius [23, 24]. It is, however, not a crucial issue in the case of the considered medical system (see Section 5), as the system requires sending short messages only, like few bytes of information, about the human body condition gathered by a liposome particle.

### 4.2. Channelrhodopsins

Channelrhodopsins, containing a retinal complex inside (see Fig. 1b), are able to create naturally occurring ion channels (pores). An important property of retinal is that, after an excitation, i.e., absorption of a photon or FRET, it opens itself, creating a pore where cations (positive ions present in human blood, like $Ca^{2+}$, $K^+$, $Na^+$) may flow through. Such a pore remains open for at least 10 ms [25], consequently, having in mind typical concentration of cations in human blood of 146 mmol/l, at least $3 \times 10^4$ cations per second flow through the channel in that time [25]. It can be measured with electrodes, as it leads to the potential change behind the layer of channelrhodopsins by 50 mV or more[3] [27]. This channelrhodopsin property means that such a molecule can serve as an *energy-voltage nano-converter*, which is extremely useful for the purpose of receiving FRET signals.

---

[2] FRET with a bioluminescent donor is also called BRET (Bioluminescent Resonance Energy Transfer).
[3] In general, the change of this potential depends on many factors, like cation type and concentration, the pool volume, electrodes capacity, etc., and can be calculated with Nernst equation [26].

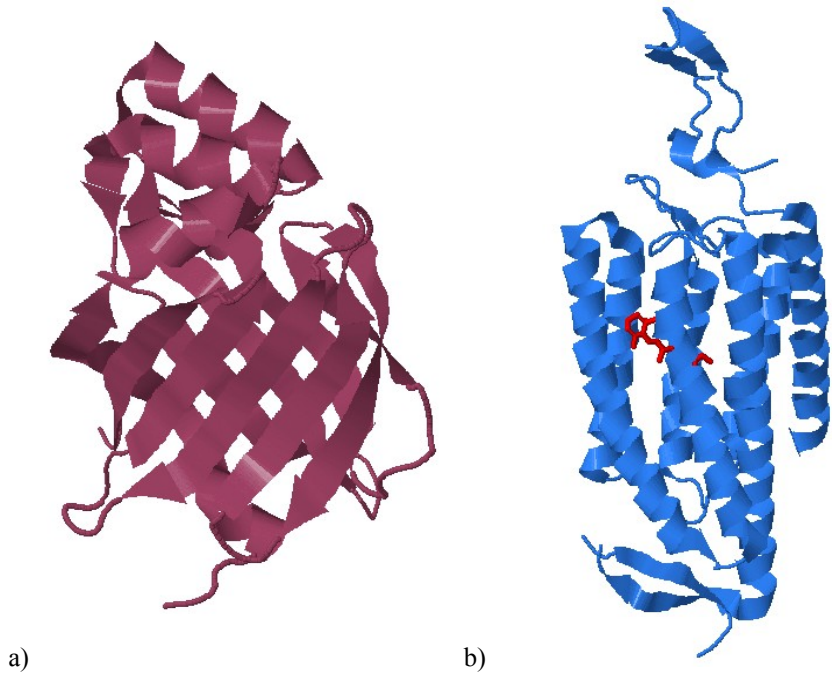

<p style="text-align:center">a)                  b)</p>

Fig. 1. Transmitter (donor) and receiver (acceptor) molecules structures derived from Protein Data Bank and visualized with Matlab (toolbox Bioinformatics): a) NanoLuc structure 5IBO [28] b) an example of channelrhodopsin structure 3UG9 [29] with retinal marked in red.

### 4.3. Luciferase-channelrhodopsin communication

Excitation of a channelrhodopsin may be realized via a photon and via FRET as well. Thus, the channelrhodopsins may work as acceptors and receive FRET signals from luciferase molecules working as donors (see Fig. 2). The FRET efficiency between a single luciferase and a single channelrhodopsin is not sufficient for telecommunication purposes and this is why it is proposed to use large matrices of molecules to communicate, exploiting the MIMO-FRET concept. As explained in Section 2, sending bit '0' is assumed to be always correct, while successful sending bit '1' requires that at least one of channelrhodopsins receives the energy from a luciferase. When the energy is properly received, channelrhodopsins open their pores, cations from the blood flow through them and the voltage inside the detector increases. Channelrhodopsins are then inactive for the time of opening the ion channels.

Bits '0' may be distinguished from bits '1' by measuring the voltage changes. The energy transfer between luciferases and channelrhodopsins is realized via FRET, but it may also happen that the channelrhodopsin layer catches a photon. These two events should be taken into account together and the probabilities of both events are considered in the performance analysis reported in Section 6.

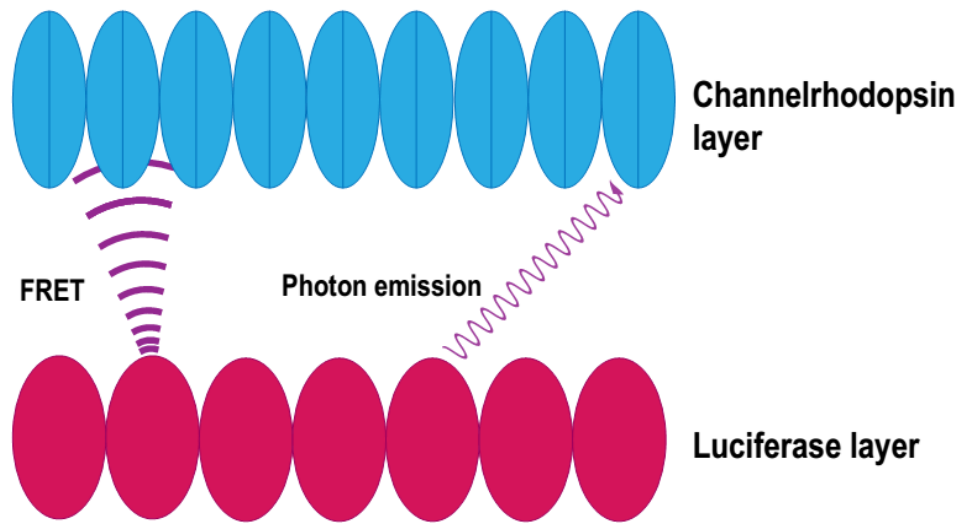

Fig. 2. The luciferase excitation energy may be transmitted to a channelrhodopsin via FRET or a photon.

After being excited and sending a signal, the luciferase molecules have also a period of inactivity, similarly as channelrhodopsins, but much longer. The luciferase turnover rate depends on the environmental conditions like temperature and substrate concentration [30]. If the analyzed communication system is located inside a human vein, we should take into account the blood parameters and its temperature[4]. Here, we assume NanoLuc luciferases that may be excited once per 0.26 s [23, 24]. This creates a bottleneck for the data transmission and this is why we suggest exciting simultaneously only a subgroup of all the luciferases when transmitting bit '1'. Other luciferases may be used during the next '1' transmission, increasing the transmission speed, i.e., the throughput of such a communication channel.

## 5. Applications in medical in-body systems

The proposed luciferase-channelrhodopsin communication, based on FRET, looks very promising for the purpose of future medical in-body systems. Inside a human body, it is rather questionable to use laser for initiating a FRET transmission and luciferases fueled by a chemical reaction are much more appropriate. Also, data collection is very straightforward if the FRET signal might be converted into the electrical one. Numerous wired detectors [31, 32], which are already used in clinical practice, might be successfully applied for measuring electrical signals. In this section, we describe an example of a medical system for gathering local data about a human body state. The system is based on liposomes circulating in human vascular system. First, we investigate the importance of liposomes in medical systems. Then, we present a medical system where liposomes collect the data inside human body and further pass it, via FRET, to a data detector.

---

[4] The luciferase turnover rate increases about 2 times per each 10 degrees Celcius of temperature increase [22].

## 5.1. Medical importance of liposomes

Almost all medical laboratory tests measuring various parameters in human blood reveal the global compound concentrations or enzyme activities in human body without spatial distribution or differentiation. In most clinical situations, these parameters are even in the whole system and such result is reliable and gives necessary information. However, in early stages of some disorders, e.g., neoplasia, due to its localized process, the global blood parameters are normal. Thus, measuring blood parameters do not allow proper identification of the disease. Local pathologies are diagnosed in medicine using various imaging techniques based on X-ray, ultrasound or radioactive nuclides (iodine, technetium) used in diagnostic procedures. Modern imaging techniques are able to reveal objects smaller than 1 cm in diameter, but are rather not used for screening patients. Also, standard X-ray or ultrasound techniques are not enough to visualize small tumors.

The local condition important from the oncological point of view is hypoxia, i.e., insufficient supply of oxygen – it may be a hallmark of tumor presence but is also extremely important in the neoplasia prognosis, i.e., tumors with hypoxia are more aggressive and resistant to therapy [33]. Hypoxia is caused by abnormalities in the vascular network and is present in majority of malignant tumors [34]. Although there is no correlation between tumor size and hypoxia [33], there are descriptions of highly hypoxic tumors with less than 1 mm size [35]. Methods for tissue oxygen detection involves several techniques and can be divided into three groups: direct measurements of oxygen in tissue, evaluation of markers produced by cells in response to hypoxia, and observation of physiological processes involved in oxygen supply [33]. It should be emphasized that only few of these techniques are approved for clinical use [33], thus the development of new techniques for hypoxia assessment in patients is required. One of the possibilities can be usage of biosensors based on liposomes.

A variety of methods for artificial lipid vesicles (liposomes) formation has been already described [36]. These methods differ depending on desired properties of the liposomes, i.e., size, the number of lipid membranes, etc.

The development of biosensors based on lipid membranes or liposome platforms has led to invention of numerous tools for environment monitoring and clinical diagnostics [36]. The lipid bilayer is the natural environment for many enzymes, antibodies, and receptors, thus it gives plenty of opportunities to build a set up for detection of various compounds. Some systems rely only on a lipid membrane without any incorporated bioelement, i.e., a measured molecule is adsorbed into a bilayer causing biochemical changes followed by conversion to an electrochemical signal (increase the trans-membrane ion current), which is simultaneously amplified [37].

Drugs acting inside cells in some point must go through the lipid membrane. This interaction with the bilayer may substantially influence drug pharmacokinetics and efficacy [38]. Modern drug delivery systems are based on polymeric coatings (chitosan or dextrans), which modify lipid-drug interactions in various ways depending on desired properties and destination of the drug. The special attention is paid to delivery of reactive plasma species into cancerous cells. High resolution structural studies revealed that synergistic effect between electric field fluctuations and lipid oxidation leads to the formation of pores. Through these structures the reactive species pass the lipid bilayer and enter into the cancer cell [39].

The long term goal of lipid vesicles studies is the development of artificial cells able to substitute function of damaged or destroyed parts of our organs. The incorporation of artificial components into a cell-like structure may not only replace the functions of the natural cell [40], but also create features not met in wild type cells, but desired from the clinical point of view.

### 5.2. Example of medical data gathering system

Here, we consider liposome particles circulating in human blood and gathering data about the human body condition and tissues pathology, e.g., tumor occurrence. There is also a wired detector, located in one of main veins, e.g., the vena cava (either of two main veins delivering deoxygenated blood to the heart), collecting the data from liposomes (see Fig. 3).

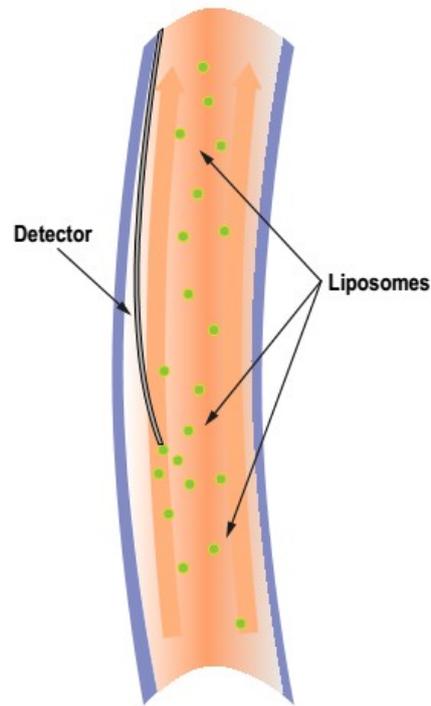

Fig. 3. The nanocommunication system located in a vein consists of a detector (a thin wire) and liposomes, tiny vesicles circulating through the human vascular system.

The liposomes work as mobile nanosensors. These particles have size up to few micrometers in diameter and they circulate in the human vascular system gathering specific data, depending on their application. At the current level of biotechnology, liposomes may be designed from basic blocks in laboratories and many components may be added inside and on their surface [41]. Liposomes can be used for diagnostic and therapeutic applications containing various markers or drugs [42, 43]. Here, we assume that a layer of luciferase molecules is attached to each liposome. A single luciferase molecule is of oblong shape, less than 10 nm large, so the layer can be built by placing luciferase molecules next to each other (Fig. 4). The whole layer, even consisting of hundreds of luciferases, is very small compared with the liposome, the size of which is on the order of few micrometers. Vesicles containing furimazine are located under the membrane with luciferase layer, to provide substrates for luciferase molecule excitations. When a liposome flows through the inferior vena cava close to the detector, an intermolecular (ligand-receptor, e.g. CD40/CD40L) bond might be created, linking the luciferases with channelrhodopsin molecules located in the detector [44]. When the bond holds, the luciferase and the channelrhodopsin layers may be fixed at a deterministic distance from each other, we assume 4 nm here, thanks to the junction system based on Annexin-V molecules (actually commercially available) [45, 46]. After establishing the bond, the data transmission via FRET is initiated, what may be realized, for example, using specific proteins called TRAFs [44]. The bond is kept as long as the data transmission occurs and is released after that via controlling the CD40/CD40L system. The liposome then returns to circulate through the blood vessels gathering data again. We assume here that there is a mechanism for attracting the liposome, floating in a vein, close to the detector. This issue is out of the scope of this paper, but it gains much attention currently, with mechanisms like magnetic field or dielectrophoresis (DEP) proposed and thoroughly studied in [47].

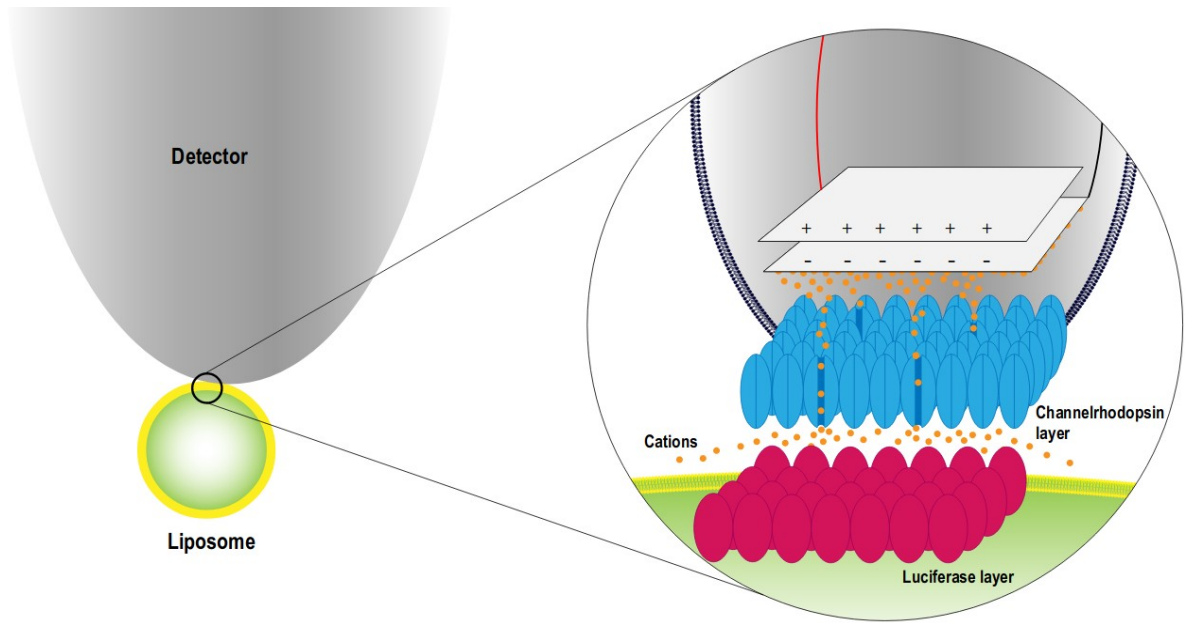

Fig. 4. The layers of luciferases and channelrhodopsins are bind with intermolecular bonds.
FRET transmission is initiated, resulting in opening pores for cations.

The detector is a device that can be mechanically put into a human large vein, like inferior vena cava. Many devices of this type exist and are popular on the medical market; the examples are vein filters and catheters used for prevention of pulmonary embolism and drug delivery [48]. Here, we assume that the top of the detector is covered with a layer of channelrhodopsin molecules, which are arranged in a squared matrix and located on a lipid membrane, using so called black lipid membrane technique [49, 50]. Above the channelrhodopsin layer (Fig. 4), there are two electrodes to measure the voltage changes caused by ions entering through channelrhodopsins.

## 6. Luciferase-channelrhodopsin communication model

The communication between the luciferases attached to a liposome (transmitter side) and channelrhodopsins at the detector (receiver side) was further modeled and validated. As explained in Section 4, information signals may be passed from a luciferase to a channelrhodopsin via FRET or photon. As both these phenomena are probabilistic ones, computer simulations were conducted.

First, we modeled both luciferase and channelrhodopsin layers using the molecular structures from Protein Data Bank (see Fig. 1). A luciferase molecule is 4.6 nm high and its diameter is 2.5 nm; we assumed having 900 luciferases arranged in a 30×30 square, therefore the size of the layer was 75×75 nm and its thickness was 4.6 nm. On the other hand, a channelrhodopsin molecule is a little larger, 5.35 nm high and 3.2 nm in diameter. The channelrhodopsin layer was planned to be wider than the luciferase one, in order to catch as much FRET and photon signals coming from luciferases as possible. Thus, we had 50×50, i.e. 2500 channelrhodopsins, so the size of this layer was 160×160 nm and it was 5.35 nm thick. Both layers were very small comparing with the dimensions of a liposome (1-10 μm) and the detector head (0.5 mm), so they could be easily mounted there. The ligand–receptor (e.g., CD40/CD40L) connection between the liposome and the detector is needed in order to create a close attachment between the luciferase and the channelrhodopsin layers; the CD40/CD40L connection is responsible for creating and disrupting the attachment [44].

Second, we modeled FRET communication between the luciferases and channelrhodopsins layers. We calculated the Förster distance $R_0$ for the luciferase–channelrhodopsin pair as 4.09 nm on the basis of their spectral characteristics [22, 51] and Eq. (2); see also Fig. 5 for the spectra of luciferase emission and channelrhodopsin absorption. Each FRET transmission was considered separately, i.e. we calculated the probability of successful transmission from the donor molecule to the particular acceptor molecule in the presence of other acceptors [3]:

$$E_{kl} = \frac{\left(\dfrac{R_0}{r_{kl}}\right)^6}{1 + R_0^6 \sum_{i=1}^{m} \dfrac{1}{r_{ki}^6}} \quad (5)$$

where $k$ and $l$ indicate the particular donor and acceptor, respectively, and $m$ is the number of acceptors. Only acceptors and donors in ground state were taken into account in the formula. These single probabilities can be summed over acceptors giving the total probability of successful transmission from $k$-th donor to any acceptor.

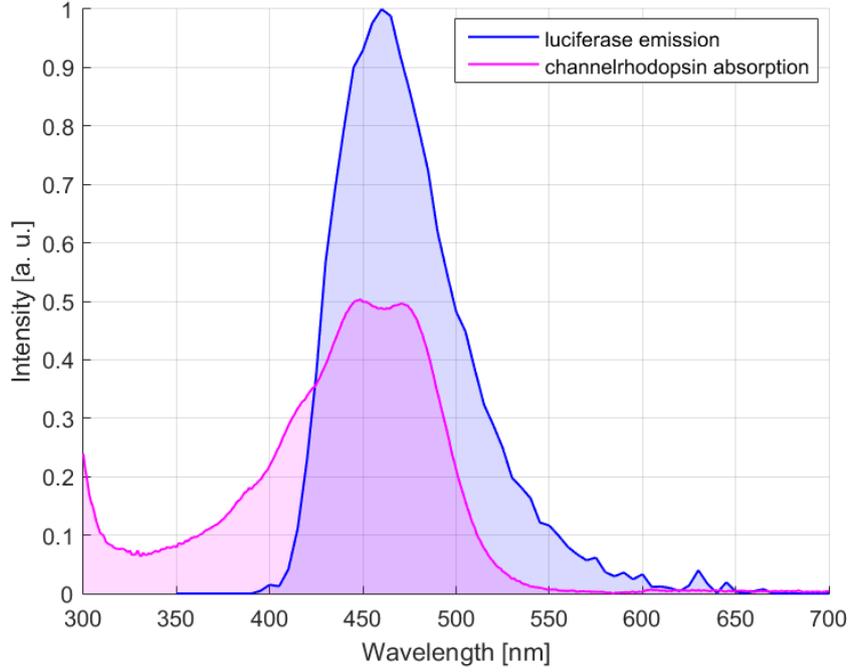

Fig. 5. Luciferase emission and channelrhodopsin absorption spectra.

Third, we also considered a possibility of the acceptor excitation via photon. The probability of hitting the retinal by a photon was calculated based on simulation of 1 million photon emissions by a single luciferase. The simulation was performed for each luciferase. We modeled retinal as a circle with the diameter of 1.35 nm, positioned in the middle of every channelrhodopsin. Direction of the emitted photon was the result of the composition of vertical and horizontal angles, which were assumed to be uniformly distributed. Keeping in mind that it can happen only when no FRET occurred, the probability of successful photon transmission between particular donor–acceptor pair can be described as:

$$E'_{kl} = 0.5 \cdot (1 - E_{kn}) \cdot p_{kl} \qquad (6)$$

where $p_{kl}$ is the probability, obtained from the simulation, of the excitation of the $l$-th channelrhodopsin by photon emitted by the $k$-th luciferase. The value $E_{kn}$ denotes the probability of FRET transmission from the $k$-th luciferase to all $n$ donors. The additional factor of 0.5 in Eq. (6) results from the fact that about 50% of photons hitting the retinal lead to its excitation [52]. Having in mind the chemical composition of blood, which was the assumed environment, no other quenching mechanisms were considered.

Having modeled both probabilities, FRET and photon excitation and adding them to each other, we finally could calculate the total probability of a successful signal transmission between the luciferase and channelrhodopsin layers.

## 7. System evaluation

Having models of luciferase and channelrhodopsin molecules and knowing the probabilities of successful transmissions, we can evaluate the performance of this communication system. We discuss it in the following subsections in respect of the bit error rate, throughput, propagation delays and energy consumption.

### 7.1. Bit error rate

The large layers of luciferases (FRET donors) and channelrhodopsins (FRET acceptors) may be used to increase the throughput, via parallel transmissions, or reliability, via simultaneous transmissions of the same bits. As a reference scenario, let us calculate the FRET efficiency in a single luciferase–channelrhodopsin communication using Eq. (1). The shortest possible donor–acceptor distance is equal to the half of the luciferase height plus the length of the ligand receptor bond plus the half of the channelrhodopsin height, which is in total 8.98 nm. The Förster distance $R_0$ for the luciferase–channelrhodopsin pair is 4.09 nm, so the FRET efficiency is only 0.89%. As explained previously, there is also a probability that a photon emitted by the luciferase hits the retinal inside the channelrhodopsin, but the probability of this event is quite minor and is equal to 0.28%. Thus, the total probability of a successful signal transmission $E_{1,1}$ equals to 1.17%. The resulting bit error rate can be calculated as $0.5(1 - E_{1,1})$ which is 49.42%. Such a value of BER is obviously not acceptable for communication purposes, which is the reason to apply the MIMO-FRET approach here, using *a fraction* of the total number of luciferases at the same time to transmit a single bit '1'. As the luciferases are characterized by quite a long time of relaxation (remaining inactive 0.26 s after sending a bit '1'), the next bit '1' is transmitted using another, equally numerous group of luciferases. We have assumed here that the number of luciferases excited when transmitting a bit '1' is controlled via regulation of furimazine (see Sections 4.1 and 5.2) admission, similarly to the mechanism present in synapses [53]. The furimazine molecules, in a chosen amount, may be packed into small synaptic vesicles (about 40 nm in size). The release of furimazine from these vesicles is triggered by calcium ions flux that, in turn, is regulated by electrical impulses that may be generated from the liposome [53]. Bits '0' are sent just by transmitting nothing, i.e., keeping the luciferases in the ground state, as it is the ON-OFF modulation.

We have analyzed 14 scenarios: using 9, 10, 12, 15, 18, 20, 25, 30, 36, 45, 50, 60, 75 and 90 out of the total number of 900 luciferases to transmit a single bit '1'. We have chosen the numbers so they are divisors of 900; in that way we could use *exactly all* the luciferases after multiple '1' transmissions. For each scenario, we have run 100 Monte Carlo simulations with $10^4$ bits ($10^6$ bits in total) using the

Matlab computing environment. The bit error rates and the respective 99% confidence intervals for all the scenarios are shown in Fig. 6. We can see that using 75 luciferases (8.3%) for a single bit '1' enables to achieve BER below $10^{-3}$, while using 90 luciferases (10%) results in BER about $10^{-4}$.

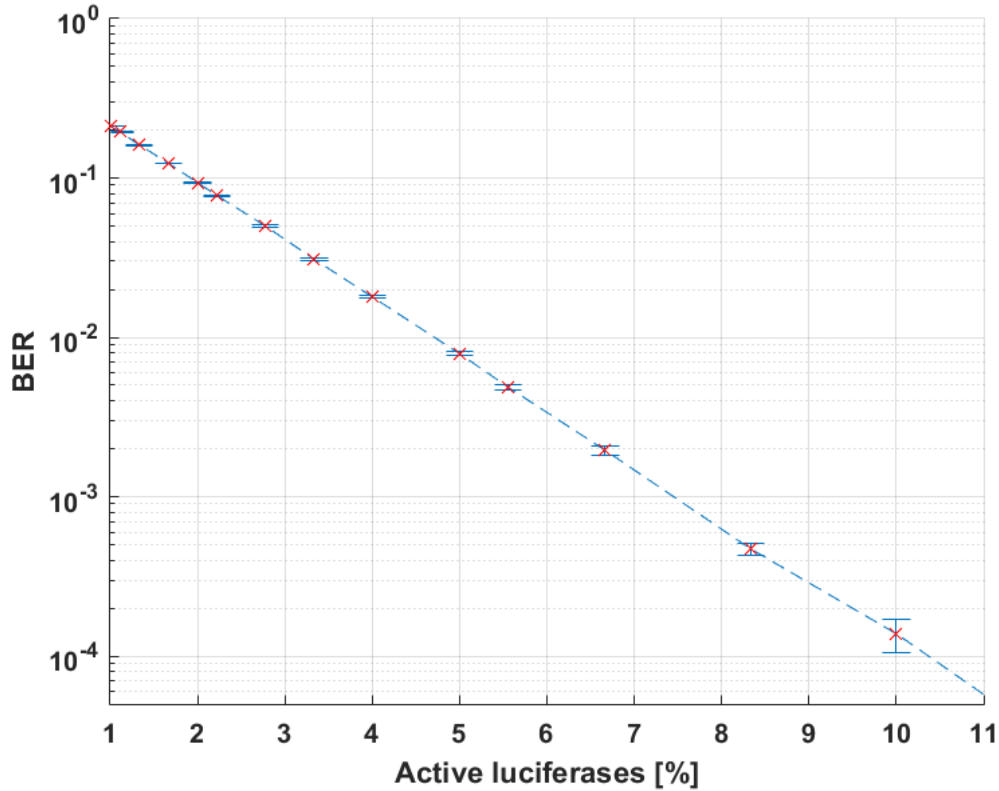

Fig. 6. Bit error rate in the liposome-detector communication as a function of the percentage of luciferases active per a single bit '1'.

## 7.2. Throughput

Having more luciferases transmitting the same bit of information increases the diversity, but, at the same time, decreases the possible communication throughput. As mentioned before, luciferases have quite a long relaxation period. For the NanoLuc type of luciferases considered in this paper, the total cycle of their activity may be approximated as 0.52 s in room temperature. Here, the value of 0.26 s is accepted, as the temperature of human blood is about 10 degrees higher (see Section 4.3). Also, this value is assumed to be constant, having in mind its very low measured variance [23]. Let us assume now that 90 luciferases, which makes 10% of their total number, transmit the same bit of information. After that transmission, these luciferases cannot be used again during 0.26 s, but we can use another 10% of luciferases for the transmission of another bit and so on, and so forth. In total, we can transmit only 10 bits in 0.26 s, so the throughput is about 38 bit/s. The throughput may be higher when fewer luciferases are involved in transmission of a single bit, i.e., the total number of 900 luciferases is divided into more groups. In Table 1, the parameters for all 14 scenarios and the respective communication throughput are given.

As mentioned before, decreasing the number of luciferases transmitting the same bit of information results in a higher throughput, but a higher BER as well. The data from Fig. 6 and Table 1 are combined together to provide Fig. 7, where the trade-off between the communication throughput and the diversity is shown. We clearly see that if BER about $10^{-3} - 10^{-4}$ is required, we can achieve the data transfer rate of 40 – 60 bit/s. This is quite a low value comparing with typical telecommunication networks; it is also a few orders of magnitude lower throughput comparing with Mbit/s values reported [4], when FRET is initiated with a laser. This throughput drop is mainly due to the long relaxation period of luciferases and comes as a price of applying these chemically induced molecules. However, it should be noted that this throughput is quite sufficient for the purpose of collecting data from liposomes in the considered medical system. A liposome, performing as a mobile nanosensor, spends about one minute to make a whole circle through the human vascular system. Spending just few more seconds when being connected to the detector, the liposome may transmit about 20 bytes which is enough to pass the gathered data. In case of specific system requirements, the sizes of luciferase and channelrhodopsin layers may be adjusted, increased or decreased, if needed.

TABLE 1

THE 14 SCENARIOS CONSIDERED IN THE NUMERICAL EXPERIMENT

| Scenario | | | | | | | | | | | | | | |
|---|---|---|---|---|---|---|---|---|---|---|---|---|---|---|
| # of luciferases transmitting a single bit | 9 | 10 | 12 | 15 | 18 | 20 | 25 | 30 | 36 | 45 | 50 | 60 | 75 | 90 |
| percentage of total number of luciferases [%] | 1 | 1.1 | 1.3 | 1.7 | 2 | 2.2 | 2.8 | 3.3 | 4 | 5 | 5.6 | 6.7 | 8.3 | 10 |
| # of luciferase groups | 100 | 90 | 75 | 60 | 50 | 45 | 36 | 30 | 25 | 20 | 18 | 15 | 12 | 10 |
| respective throughput [bits/s] | 385 | 346 | 288 | 231 | 192 | 173 | 138 | 115 | 96 | 77 | 69 | 58 | 46 | 38 |

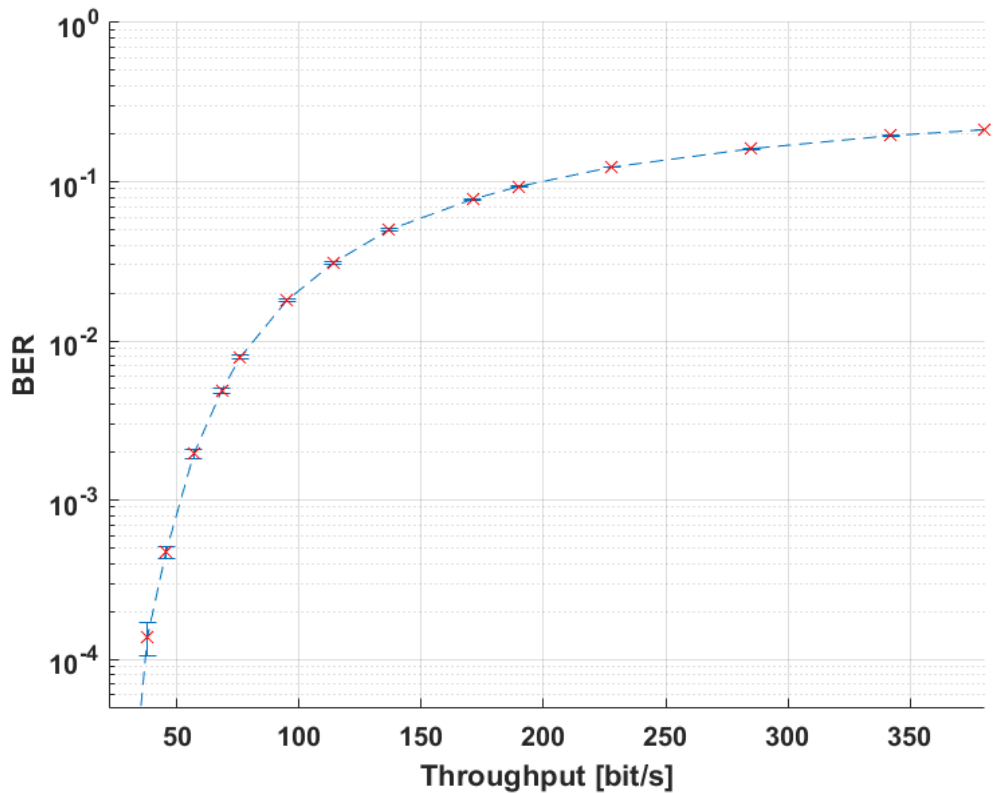

Fig. 7. Bit error rate in the liposome-detector communication as a function of achieved throughput.

### 7.3. Propagation delay

In order to assess how fast the signal propagates from a luciferase to a channelrhodopsin, two factors should be taken into account. The propagation time is very short in both cases of a photon emission and a FRET. In the former, the photon travels with the speed of light, so the distance of a few nanometers is reached in a fraction of a femtosecond. The photon absorption itself is a very fast process as well, also taking less than $10^{-15}$ [17]. In the latter, FRET is realized via a dipole-dipole interaction of a donor and an acceptor, and it is assumed that its delay is insignificant. However, before the energy transfer or the photon emission occurs, the luciferase keeps the energy for a certain period of time called its excitation lifetime. Its average value is about 5-10 nanoseconds [54], but its actual value might considerably vary, as this lifetime is given by the exponential distribution.

### 7.4. Energy consumption

Finally, we can calculate the energy transferred and dissipated during data transmission in the considered system. The emission energy of a single NanoLuc luciferase (FRET or a photon) can be assessed, according to the Planck–Einstein relation (energy = Planck constant × frequency, where all frequencies are averaged over emission intensities), as $3.8 \times 10^{-19}$ J. Transmitting bit „0" costs no energy, while transmitting bit „1" (assuming 10% of luciferases are excited, BER = $10^{-4}$, see above) means exciting 90 luciferases molecules, i.e., costs ca. $3.4 \times 10^{-17}$ J. Assuming equal number of "0" and „1" bits, we may assess that, on average, during transmission of a single bit, we have $1.7 \times 10^{-17}$ J transmitted. Additionally, taking into account the efficiency of the luciferase reaction being about 28% [23], we have $4.4 \times 10^{-17}$ J dissipated, so the total energy related to the transmission of a single bit is about $6.1 \times 10^{-17}$ J. This energy is lower than in case of other molecular communication mechanisms, where energy per bit is between $10^{-16}$ and $10^{-13}$ J (see Section 2).

## 8. Conclusions and open issues

In this paper, we propose application of two molecule types, luciferases and channelrhodopsins, to solve the following two important issues in the FRET-based nanocommunications: signal generation and its storage. Typically, FRET communication is initiated with a laser, but here, we suggest using a chemical reaction when a luciferase initiates FRET with energy provided by ATP. On the receiver side of the communication system, we propose channelrhodopsins which can convert FRET signals into electrical potential. We discuss medical in-body systems designed for gathering data on human body condition where such FRET-based communication might be very useful. We consider a system with liposomes circulating in human vascular system and collecting data about local tissue pathologies, e.g., tumor occurrence. The collected data is further forwarded to a detector, placed into a vein. As there is a layer of luciferases on each liposome and the detector is covered with a layer of channelrhodopsins, the data transmission is realized via FRET in the luciferase-channelrhodopsin interface. The communication is validated with numerical calculations and we provide the corresponding throughput, bit error rate, propagation delay and energy consumption.

Still, there are some aspects of the system which, at the current level of technology, are not solved and require further studies. The first aspect concerns *writing and reading the collected data* in a liposome. Storing data in DNA could be an option and it is currently quite a popular research topic [55]. Data storage in DNA was recently presented [56]; the main advantage of this solution is its extremely high efficiency in terms of the occupied space: 1.98 bits per nucleotide. The potential drawback is a relatively low reading rate compared to commonly used computer systems. Additionally, reading information stored in DNA needs special enzymes and reaction conditions *in vitro*. Thus, it is an open problem how to organize data storage and reading in a simple way suitable for the nanocommunication system presented above. Moreover, reading the collected data should be associated with *controlling and synchronizing the chemical reaction exciting luciferase molecules*, as it initiates data transmission. Another issue is a mechanism of *attracting liposomes by the detector*. The electric field is rather not a solution, as it might influence other particles in human blood. Instead, the magnetic field or dielectrophoresis effect might be used [47, 57, 58]. Finally, there is a question regarding the *throughput* achievable in the liposome–channelrhodopsin communication. With NanoLuc luciferases considered in the paper, the throughput is about 40 bit/s, which is sufficient for systems focused on collecting medical data inside a human body. There are still few options to increase the throughput, if needed. First, in some scenarios, the system temperature might be slightly increased, what in turn improves the luciferase turnover rate [24]. Second, the luciferase and channelrhodopsin molecules might be carefully positioned in their respective membranes, to keep the orientations of their dipoles in parallel, increasing in that way the $\kappa^2$ parameter in Eq. (2) from 0.476 assumed here to even 4 (more than 8 times). Third, the layers of luciferases and channelrhodopsin might be put even closer than 4 nm from each other; it depends on the Annexin-V molecules used to mount luciferases on the liposome membrane [46]. If, however, an even higher throughput is required, one can consider using fluorescent dyes (e.g., Alexa, Atto or DyLight) as donors at the liposome excited by the light provided by an optical fiber inside the detector. Fluorescent dyes are able to emit photons or FRET signals with much higher frequency, supporting data throughput of Mbit/s [4]. Nonetheless, in this case, reading the data collected in the liposome should be realized by modulation of the light coming from the detector, which is a task still not solved.

## Acknowledgements

This work was supported by the Polish Ministry of Science and Higher Education with the subvention funds of the Faculty of Computer Science, Electronics and Telecommunications of AGH University of Science and Technology. It is also a result of the research project DEC-2013/11/N/NZ6/02003 financed by the National Science Centre. It was also supported in part by the PLGrid Infrastructure.